%% file: Albuquerque.tex
\documentclass[final,3p,times,twocolumn]{elsarticle}
 \biboptions{comma,sort&compress}
\usepackage{here}
 \usepackage{graphicx}
  \usepackage{epsfig}

\def\beq{\begin{equation}}
\def\eeq{\end{equation}}
\def\bea{\begin{eqnarray}}
\def\eea{\end{eqnarray}}
\def\bq{\begin{quote}}
\def\eq{\end{quote}}

\def\nnb{\nonumber}
\def\ga{\left(}
\def\dr{\right)}

\def\rar{\rightarrow}

\def\nnb{\nonumber}
\def\la{\langle}
\def\ra{\rangle}
\def\nin{\noindent}
\def\ba{\vspace*{-0.2cm}\begin{array}}
\def\ea{\end{array}\vspace*{-0.2cm}}

\def\b{$\bullet~$}

\def\als{\alpha_s}

\def\gg2{ \la\alpha_s G^2 \ra}
\def\gg3{g^3f_{abc}\la G^aG^bG^c \ra}
\def\ggg4{\la\als^2G^4\ra}

\journal{Nuc. Phys. (Proc. Suppl.)}

\begin{document}

\begin{frontmatter}



\title{Doubly heavy Baryons from QCD Spectral Sum Rules}

\author[label1,label2]{R.M. Albuquerque\corref{cor1}\fnref{label3}}
  \address[label1]{Instituto de F\'{\i}sica, Universidade de S\~{a}o Paulo, 
  C.P. 66318, 05389-970 S\~{a}o Paulo, SP, Brazil}
  \cortext[cor1]{Speaker}
  \fntext[label3]{FAPESP thesis fellow within the France-Brazil bilateral exchange program.}
  \ead{rma@if.usp.br}

\author[label2]{S. Narison}
  \address[label2]{Laboratoire de Physique Th\'eorique et 
  Astroparticules, CNRS-IN2P3 \& Universit\'e de Montpellier II, \\
  Case 070, Place Eug\`ene Bataillon, 34095 - Montpellier Cedex 05, France.}
  \ead{snarison@yahoo.fr}


\begin{abstract}
\noindent
We consider the ratios of  doubly heavy baryon masses using Double Ratios of Sum Rules (DRSR), 
which are more accurate than the usual simple ratios used for getting hadron 
masses. Our results are comparable with the ones from potential models. In our approach, the $\alpha_s$ corrections induced by the anomalous dimensions of the correlators are the main sources of the $\Xi^*_{QQ}- \Xi_{QQ}$ mass-splittings, which seem to indicate a $1/M_Q$ behaviour and can only allow the electromagnetic decay $\Xi^*_{QQ}\to\Xi_{QQ}+\gamma$ but not to ${\Xi_{QQ }}+ \pi$. Our results also show that the SU(3) mass-splittings are (almost) independent of the spin of the baryons and behave approximately like $1/M_Q$, which could be understood from the QCD expressions of the corresponding two-point correlator. Our results  can improved by including radiative corrections to the SU(3) breaking terms and can be tested, in the near future, at Tevatron and LHCb.
\end{abstract}

\begin{keyword}
QCD spectral sum rules, baryon spectroscopy, heavy quarks.

\end{keyword}
\end{frontmatter}

\section{Introduction}
\nin
In a previous paper \cite{HBARYON}, we have considered, using Double Ratios (DRSR)
\,\cite{SNGh} of QCD spectral sum rules (QSSR) \cite{SVZ,SNB}, 
the splittings due to SU$(3)$ breakings of the baryons made with one heavy quark. This project has been pursued in the case of doubly heavy baryons in \cite{RN}, which will be reviewed in this talk. \\
The absolute values of the doubly heavy baryon masses of spin 1/2 $(\Xi_{QQ}\equiv QQu)$ and 
spin 3/2 $(\Xi^*_{QQ}\equiv QQu)$ have been obtained using QCD spectral sum rules (QSSR) 
(for the first time) in \cite{BAGAN} with the results in GeV:
\bea
M_{\Xi^*_{cc}}= 3.58(5)~&,&~~M_{\Xi^*_{bb}}=10.33(1.09)~,\nnb\\
M_{\Xi_{cc}}=3.48(6)~&,&~~M_{\Xi_{bb}}=9.94(91)~,
\label{eq:bagan}
\eea
and in \cite{BC}:
\bea
M_{\Xi_{bcu}}&=&6.86(28)~.
\label{eq:bc}
\eea
More recently \cite{ZHANG,WANG}, some results have been obtained using some particular choices of the interpolating currents. The predictions for $M_{\Xi^*_{cc}}$ and $M_{\Xi_{cc}}$ are in good agreement with the experimental 
candidate $M_{\Xi_{cc}}=3518.9$ MeV  \cite{PDG}. In the following, we shall improve these previous predictions using the DRSR for estimating the ratio of the $3/2$ over the $1/2$ baryon masses as well as their splittings due to SU(3) breakings, which we shall compare with some potential model predictions \cite{BC,RICHARD,BRAC,VIJANDE}. 

\section{The Interpolating Currents}
\nin
\b {\bf For the spin 1/2  \boldmath$(QQq)$ baryons },  and following Ref. \cite{BAGAN}, we work with the 
lowest dimension currents:
\beq
J_{\Xi_Q}=\epsilon_{\alpha\beta\lambda}\left[(Q_\alpha^TC\gamma_5q_\beta)+
b(Q_\alpha^TCq_\beta)\gamma_5\right] Q_\lambda, \label{cur1}
\eeq
where $q\equiv d,s$ are light quark fields, $Q\equiv c,b$ are heavy quark fields, $b$ is {\it a priori} an 
arbitrary mixing parameter. Its value has been found to be: $b=-1/5$, 
in the case of light baryons \cite{JAMI2} and in the range \cite{BAGAN1,BAGAN2,HBARYON}:
\beq
-0.5\leq b\leq 0.5~,
\label{eq:mixing2}
\eeq
for non-strange heavy baryons. The corresponding two-point correlator reads:
\bea
S(q)&=&i\int d^4x~ e^{iqx}~ \la 0\vert {\cal T} \overline{J}_{\Xi_Q}(x)J_{\Xi_Q}(0)\vert 0\ra\nnb\\
&\equiv& \hat q F_1  +F_2~,
\eea
where $F_1$ and $F_2$ are two invariant functions. \\

\b {\bf For the spin 3/2   \boldmath$(QQq)$ baryons}, we  also follow  Ref. \cite{BAGAN} and work with the 
interpolating currents:
\beq
J^\mu_{\Xi^*_Q}=
\sqrt{1\over3}\epsilon_{\alpha\beta\lambda}\big{[}
2(Q_\alpha^TC\gamma^\mu d_\beta)Q_\lambda+(Q_\alpha^TC\gamma^\mu Q_\beta)q_\lambda\big{]}
\label{cur}
\eeq
The corresponding two-point correlator reads:
\bea
S^{\mu\nu}(q)&=&i\int d^4x~ e^{iqx}~ \la 0\vert {\cal T} \overline{J}^\mu_{\Xi^*_Q}(x)J^\nu_{\Xi^*_Q}(0)\vert 0\ra\nnb\\
&\equiv& g^{\mu\nu}\ga \hat q F_1  +F_2\dr+\dots~ 
\eea

\section{The  Two-Point Correlator in QCD}
\nin
The expressions of the two-point correlator using the previous interpolating currents have been obtained in 
the chiral limit $m_q=0$ and including the mixed condensate contributions by \cite{BAGAN}. In this work, 
we extend these results by including the linear strange quark mass corrections to the perturbative and 
$\la\bar ss\ra$ condensate contributions. The explicit expressions can be found in Ref. \cite{RN}.


\vspace*{-0.5cm}
\section{Form of the Sum Rules}
\label{sec:qssr}
\vspace*{-0.25cm}
 \nin
We parametrize the spectral function using the standard duality ansatz: ``one resonance"+ ``QCD continuum". 
The QCD continuum starts from a threshold $t_c$ and comes from the discontinuity of the QCD diagrams. 
Transferring its contribution to the QCD side of the sum rule, one obtains the finite energy inverse Laplace 
sum rules \cite{SVZ,BB,SNDR}. Consistently, we also take into account the SU$(3)$ breaking at the 
continuum threshold $t_c$:
\bea
  \sqrt{t_c}\vert_{SU(3)}&\simeq&   \ga\sqrt{t_c}\vert_{SU(2)}\equiv \sqrt{t_c}\dr + \bar m_s~,
\eea
where $\bar m_{s}$  is the running strange quark mass. As we do an expansion in $m_s$, we take the threshold 
$t_q=4m_Q^2$ for consistency, where $m_Q$ is the heavy quark mass, which we shall take in the range covered by 
the running and on-shell mass because of its ambiguous definition  when working 
to lowest order (LO).  As usually done in the sum rule literature, one can estimate the baryon masses from the following ratios
$(i=1,2)$:
{\small
\bea \hspace{-0.6cm}
{\cal R}^q_i={\int_{t_q}^{t_c}dt~t~
e^{-t\tau}~{\rm Im}F_{i}(t)\over \int_{t_q}^{t_c}dt~
e^{-t\tau}~{\rm Im}F_{i}(t)}~,
&& \hspace{-0.2cm}
{\cal R}^q_{21}={\int_{t_q}^{t_c}dt~
e^{-t\tau}~{\rm Im}F_{2}(t)\over \int_{t_q}^{t_c}dt~
e^{-t\tau}~{\rm Im}F_{1}(t)}
\label{eq:ratio}
\eea}
where at the $\tau$-stability point :
\beq
M_{B^{(*)}_q}\simeq \sqrt{{\cal R}^q_i}\simeq {\cal R}^q_{21}~,~~~~~(i=1,2)~.
\eeq
These predictions lead to a  typical uncertainty of 10-15\% \cite{BAGAN,BAGAN2,BC}, which are not competitive
compared with predictions from some other approaches, especially from potential models \cite{BC,RICHARD}. 
In order to improve the QSSR predictions, we work with the double ratio of sum rules (DRSR):
\beq
r^{sd}_i\equiv \sqrt{{\cal R}^s_i\over {\cal R}^d_i}~,~~~(i=1,2)~;~~~~~~~~~~r^{sd}_{21}\equiv {{\cal R}^s_{21}\over {\cal R}^d_{21}}~,
\label{eq:2ratio}
\eeq
which take directly into account the SU$(3)$ breaking effects. These quantities are obviously less sensitive to the 
choice of the heavy quark masses and to the value of the $t_c$ than the simple ratios ${\cal R}_i$ 
and  ${\cal R}_{21}$. 

\section{The  \boldmath $\Xi^*_{QQ}/\Xi_{QQ }$ mass ratio}
  \nin
 We extract the mass ratios using the DRSR analogue of the one in Eq. (\ref{eq:2ratio}) which we denote by:
 \beq
r^{3/1}_i\equiv \sqrt{{\cal R}^3_i\over {\cal R}^1_i}~,~~~(i=1,2)~;~~~~~~~~r^{3/1}_{21}\equiv {{\cal R}^3_{21}\over {\cal R}^1_{21}}~,
\label{eq:2ratio13}
\eeq
where the upper indices 3 and 1 correspond respectively to the spin 3/2 and 1/2 channels. We use the QCD 
expressions of the two-point correlators given by \cite{BAGAN} which we have checked. In our analysis, we 
truncate the QCD series  at the dimension 4 condensates until which we have calculated the $m_s$ corrections. 
We shall only include the effect of the mixed condensate (if necessary) for controlling the accuracy of the approach or for 
improving the $\tau$ or/and $t_c$ stability of the analysis. 

\subsection*{\b The charm quark channel}
\begin{figure}[t]
\begin{center}
\includegraphics[width=4.6cm]{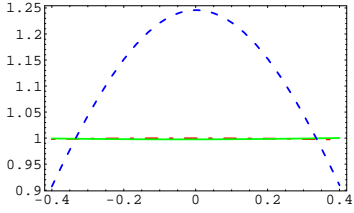}
\vspace*{-0.3cm}
\caption{\footnotesize  {\bf Charm quark :}  $b$-behaviour of the different DRSR given $\tau=0.8$ GeV$^{-2}$ and $t_c=25$ GeV$^2$.  $r^{3/1}_1$ dot-dashed line (red); $r^{3/1}_2$ continuous line (green); $r^{3/1}_{12}$ dashed line (blue). 
We have used $m_c=1.26$ GeV and the other QCD parameters in ref. \cite{RN}. }
\label{fig:r13b}
\end{center}
\vspace*{-0.5cm}
\end{figure}

\begin{figure}[t]
\begin{center}
\includegraphics[width=4.6cm]{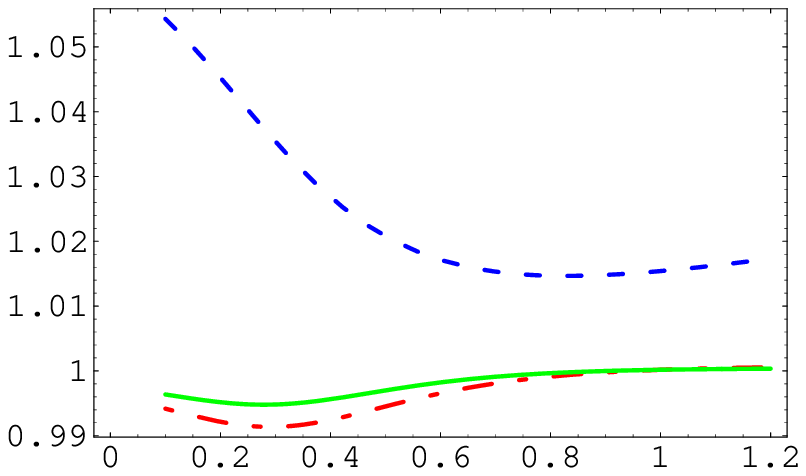}
\includegraphics[width=4.6cm]{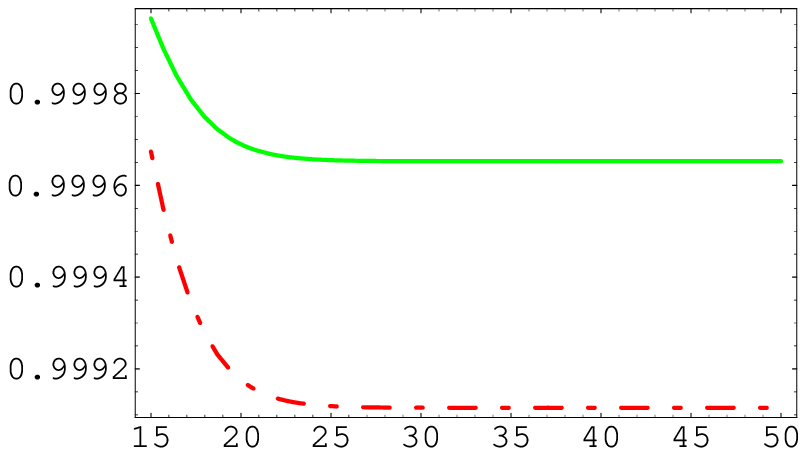}
\vspace*{-0.3cm}
\caption{\footnotesize {\boldmath ${\Xi^*_{cc}}/ {\Xi_{cc }}$}
a): $\tau$-behaviour of $r^{3/1}_1$: dot-dashed line (red) and $r^{3/1}_2$ : continuous line (green) with $b=-0.35$ and $t_c=25$ GeV$^{2}$. b): $t_c$-behaviour of $r^{3/1}_1$: dot-dashed line (red) and $r^{3/1}_2$ : continuous line (green) with $b=-0.35$ and $\tau=0.8$ GeV$^{-2}$.}
\label{fig:r13tc}
\end{center}
\vspace*{-0.5cm}
\end{figure}

\nin
Fixing $t_c=25$ GeV$^2$ and $\tau=$ 0.8 GeV$^{-2}$, which are inside the $t_c$ and $\tau$-stability regions 
(see Fig. \ref{fig:r13tc}a and  Fig. \ref{fig:r13tc}b), we show in Fig. \ref{fig:r13b} the $b$-behaviour of $r^{3/1}$ which 
shows that $r^{3/1}_1$ and $r^{3/1}_2$ are very stable but not $r^{3/1}_{12}$. We then eliminate $r^{3/1}_{12}$, 
where one can notice some common 
solutions for:
\beq
b\simeq -0.35~,~~~~~~~{\rm and}~~~~~~~b\simeq +0.2~,
\eeq
which are inside the range given in Eq. (\ref{eq:mixing2}). For definiteness, we fix $b=-0.35$ and study the 
$\tau$-dependence of the result in Fig. \ref{fig:r13tc}a and its $t_c$-dependence  in Fig. \ref{fig:r13tc}b.

\nin
The large stability in $t_c$ confirms our expectation for the weak $t_c$-dependence of the DRSR. In these 
figures, we have used $m_c=1.26$ GeV and have checked that the results are insensitive to the change of 
mass to $m_c=1.47$ GeV. We have also checked that the inclusion of the mixed condensate contribution 
does not affect the present result (within the high-accuracy obtained here) obtained  by retaining the 
dimension-4 condensates. 
  \subsection*{\b Results of the analysis}
\nin
From the analysis of the charm and bottom quark channels, we deduce with high accuracy to lowest order:
${M_{\Xi^*_{cc}}/ M_{\Xi_{cc }}}= 0.9994(3)$ {\rm and} ${M_{\Xi^*_{bb }}/ M_{\Xi_{bb}}}= 1.0000$,
while the inclusion of the radiative corrections  induced by the anomalous dimension of the correlators modifies the previous results to (see Table \ref{tab:summary}):
\beq
{M_{\Xi^*_{cc}}\over M_{\Xi_{cc }}}= 1.0167(19)~,~{M_{\Xi^*_{bb }}\over M_{\Xi_{bb}}}= 1.0019(3)~,
\label{eq:chicc}
\eeq
which would correspond to the mass-splittings in MeV:
\beq
{M_{\Xi^*_{cc}}- M_{\Xi_{cc }}}= 59(7)~,~{M_{\Xi^*_{bb }}- M_{\Xi_{bb}}}=19(3)~,
\label{eq:chibb}
\eeq
 comparable with standard potential models \cite{BC,RICHARD} but not with the one of about  24 MeV obtained in 
\cite{VIJANDE} for the charm (see Table~\ref{tab:summary}). Our result excludes the possibility that $M_{\Xi^*_{QQ}}\geq M_{\Xi_Q}+m_\pi$, indicating that the $\Xi^*_{QQ}$  can only decay electromagnetically but not to ${\Xi_Q}+\pi$.
\section{The \boldmath $\Omega_{QQ}/\Xi_{QQ }$  mass ratio}
\begin{figure}[hbt]
\begin{center}
\includegraphics[width=4.6cm]{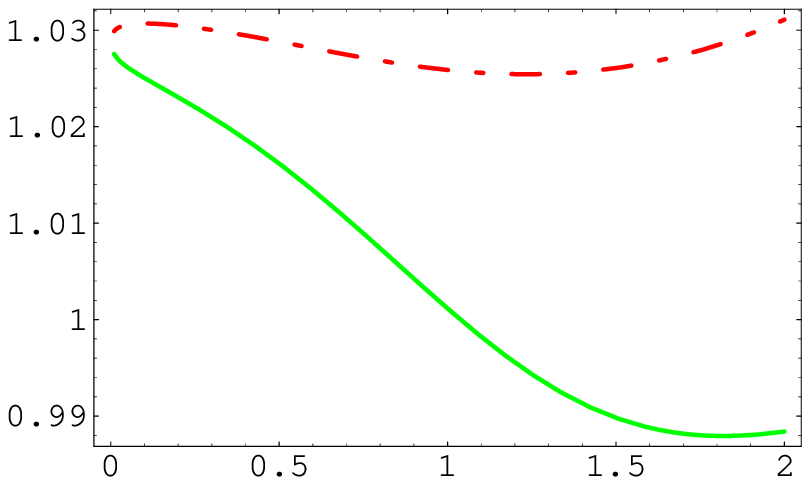}
\includegraphics[width=4.6cm]{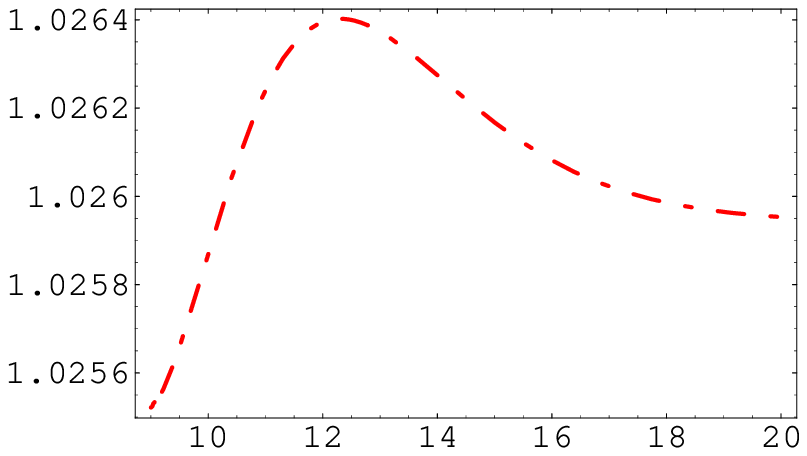}
\vspace*{-0.3cm}
\caption{\footnotesize {\boldmath$\Omega_{cc}/\Xi_{cc}$} a): $\tau$-behaviour of $r^{sd}_2(cc)$: continuous line (green) and $r^{sd}_1(cc)$: dot-dashed line (red) in the charm quark channel for $b=-0.35$, $t_c=12$ GeV$^{2}$ and $m_c=1.26$ GeV. 
b):  $t_c$-behaviour of $r^{sd}_1(cc)$ for $\tau=1$ GeV$^{-2}$: dot-dashed line (red) }
\label{fig:chic_tau}
\end{center}
\end{figure}

\nin
We use the DRSR in Eq. (\ref{eq:2ratio}) where their QCD expressions 
can be obtained from the one of 
the two-point correlator 
in \cite{BAGAN,BC}, while the new quark mass corrections can be found in \cite{RN}. The analysis for the charm quark is shown in Fig.~\ref{fig:chic_tau}, from which we can deduce the result given in Table~\ref{tab:summary}. 
\nin
A similar analysis  for the bottom quark is also given in Table 
\ref{tab:summary}.  We deduce from the ratios (in units of MeV):
\beq
M_{\Omega_{cc}}- M_{\Xi_{cc}}= 92(24)~,~~ M_{\Omega_{bb}}- M_{\Xi_{bb}}\simeq 49(13)~.
\label{eq:omegac}
\eeq
Our results indicate an approximate decrease like $1/m_Q$ of the mass splittings from the $c$ to the 
$b$ quark channels. This behaviour can be qualitatively understood from the QCD expressions of the 
corresponding correlator, where the $m_s$ corrections enter like $m_s/m_Q$, and which  can be checked 
using alternative methods.

\section{The  \boldmath $\Omega^*_{QQ}/\Xi^*_{QQ}$ mass ratio}
\nin
We pursue our analysis for the spin 3/2 baryons. We deduce, for the charm quark, at the stability regions,
the ratios given in Table \ref{tab:summary} and the corresponding mass-splittings (in units of MeV):
\beq
 M_{\Omega^*_{cc}}- M_{\Xi^*_{cc}}= 94(27) ~,~~M_{\Omega^*_{bb}}- M_{\Xi^*_{bb}}= 50(15)~,
\label{eq:omega*c}
\eeq
which agree with the potential model results given in \cite{BC} (see Table \ref{tab:summary}). 
Again, like in the case of spin 1/2 
baryons, the SU(3) mass-differences appear to behave like  $1/M_Q$, which can 
be inspected from the QCD expressions of the two-point correlator. \\
One can also observe that the mass-splittings are almost the same for the spin 1/2 and  3/2 baryons.

\section{The  \boldmath $\Omega_{bc}/\Xi_{bc}$ mass ratio}
\nin
The $\Xi_{bc}$ and the $\Omega_{bc}$ spin 1/2 baryons can be described by the corresponding currents \cite{BAGAN,BC}:
\bea
J_{\Lambda_{bc}}&=&\epsilon_{\alpha\beta\lambda}\left[(c_\alpha^TC\gamma_5d_\beta)+k(c_\alpha^TCd_\beta)\gamma_5\right]
b_\lambda, \nnb\\
J_{\Omega_{bc}}&=&J_{\Xi_{bc}}  ~~~(d\rar s)~,
\label{curl}
\eea
where $d,s$ are light quark fields, $c,b$ are heavy quark fields and $k$ is {\it a priori} an arbitrary mixing 
parameter. Like in previous sections, we study the different ratio of moments for this case. The  $b$-stability is obtained for $k\pm 0.05$ while the $\tau$ and $t_c$ 
behaviours are also very stable at which we deduce the DRSR in Table \ref{tab:summary} and the corresponding splitting:
\beq
 M_{\Omega_{bc}}- M_{\Xi_{bc}}= 41(7) ~{\rm MeV}~.
\label{eq:lambdamass}
\eeq
{\scriptsize
\begin{table}[hbt]
\setlength{\tabcolsep}{0.15pc}
 \caption{\scriptsize    QSSR predictions for the doubly heavy baryons mass ratios and splittings, which we compare with the Potential Model (PM) range of results in \cite{BC,BRAC}.  The PM prediction for the spin 3/2 is an average with the one for spin 1/2. The mass inputs are in GeV and the mass-splittings are in MeV.}
    {\footnotesize
\begin{tabular}{llll}
&\\
\hline
Mass ratios&Mass inputs&Mass plittings& PM   \\
\hline
${{\Xi^*_{cc}}/ {\Xi_{cc }}}= 1.0167(19)$&${\Xi_{cc }}=3.52$\cite{PDG}& ${{\Xi^*_{cc }}- {\Xi_{cc}}}=59(7)$&70-93\\
${{\Xi^*_{bb }}/ {\Xi_{bb}}}=1.0019(3)$&${\Xi_{bb}}=9.94$\cite{BAGAN}&${{\Xi^*_{bb }}- {\Xi_{bb}}}= 19(3)$&30-38\\
${{\Omega_{cc}}/ {\Xi_{cc}}}=1.0260(70)$&${\Xi_{cc }}=3.52$\cite{PDG}& ${\Omega_{cc}}- {\Xi_{cc}}= 92(24) $&90-102\\
${\Omega_{bb}}/ {\Xi_{bb}}=1.0049(13)$&${\Xi_{bb}}=9.94$\cite{BAGAN}&${\Omega_{bb}}- {\Xi_{bb}}= 49(13)$&60-73\\
${{\Omega^*_{cc}}/ {\Xi^*_{cc}}}=1.0260(75)$&${\Xi^*_{cc }}=3.58^{ ~*)}$&$ {\Omega^*_{cc}}- {\Xi^*_{cc}}= 94(27)$&91-100\\
${{\Omega^*_{bb}}/ {\Xi^*_{bb}}}=1.0050(15)$&${\Xi^*_{bb}}=9.96^{~ *)}$&${\Omega^*_{bb}}- {\Xi^*_{bb}}= 50(15)$&60-72\\
${{\Omega_{bc}}/ {\Xi_{bc}}}=1.0060(17)$&${\Xi_{bc}}=6.86$\cite{BC}&${\Omega_{bc}}- {\Xi_{bc}}= 41(7)$&70-89\\
\hline
\end{tabular}
\begin{quote}
\scriptsize$^{ *)}$ We have combined your results for the mass-splittings with the experimental value of 
$M_{\Xi_{cc}}$ and with the central value of $M_{\Xi_{bb}}$ in Eq. (\ref{eq:bagan}). 
\end{quote} 
}
\label{tab:summary}
\end{table}
}
\vspace*{-0.5cm}
\nin
\section{Conclusions}
\nin
Our different results are summarized in Table \ref{tab:summary} and agree in most cases with the potential model predictions given in \cite{BC,RICHARD}:\\
-- The mass-splittings between the spin 3/2 and 1/2 baryons, derived in Eqs. (\ref{eq:chicc}) and (\ref{eq:chibb}) are essentially due to the radiative corrections induced by the anomalous dimensions of the two-point correlator and seems to behave like $1/M_Q$. \\
-- For the SU(3) mass-splittings, our results derived in Eq. (\ref{eq:omegac}) for the spin 1/2  and in Eq. (\ref{eq:omega*c}) for the spin 3/2 indicate that the splittings due to the SU(3) breaking are almost independent on the spin of the heavy baryons and approximately behave like $1/M_Q$. These mass-behaviours can  be qualitatively understood from the QCD expressions of the corresponding correlators where the leading mass corrections behave like $m_s/m_Q$.  \\
-- Finally, we obtain, in Eq. (\ref{eq:lambdamass}), the SU(3) mass-splittings between the $\Omega(bcs)$ and $\Xi(bcd)$, which is about 1/2 of the potential model prediction. \\
  Our previous predictions can improved by including radiative corrections to the SU(3) breaking terms and can be tested, in the near future, at Tevatron and LHCb.

\section*{Acknowledgements}
\nin
We thank M. Nielsen, J-M. Richard and V.I. Zakharov for some discussions. R.M.A acknowledges the LPTA-Montpellier 
for the hospitality where this work has been done. This work has been partly supported by the CNRS-IN2P3 within the 
Non-perturbative QCD program in Hadron Physics.

\input{bib_sample}

\end{document}